# Impact of Load Models on Power Flow Optimization

Marko Jereminov[1], Bryan Hooi[2], Amritanshu Pandey[1], Hyun-Ah Song[2], Christos Faloutsos[2], Larry Pileggi[1]

[1]Dept. of Electrical and Computer Engineering
Carnegie Mellon University
Pittsburgh, PA

[2]School of Computer Science
Carnegie Mellon University
Pittsburgh, PA

*Abstract*—Aggregated load models, such as PQ and ZIP, are used to represent the approximated load demand at specific buses in grid simulation and optimization problems. In this paper we examine the impact of model choice on the optimal power flow solution and demonstrate that it is possible for different load models to represent the same amount of real and reactive power at the optimal solution yet correspond to completely different grid operating points. We introduce the metric derived from the maximum power transfer theorem to identify the behavior of an aggregated model in the OPF formulation to indicate its possible limitations. A dataset from the Carnegie Mellon campus is used to characterize three types of load models using a time-series machine learning algorithm, from which the optimal power flow results demonstrate that the choice of load model type has a significant impact on the solution set points. For example, our results show that the PQ load accurately characterizes the CMU data behavior correctly for only 16.7% of the cases.

*Index Terms*—BIG load, circuit formalism, circuit theory, equivalent circuit, power grid optimization, optimal power flow, split-circuit formulation, ZIP load model.

## I. INTRODUCTION

Electricity market and power grid analyses represent the key component in maintaining the grid operation efficiently and reliably. Therefore, in addition to robust power grid simulation and optimization algorithms, it is of utmost importance to develop the modeling methodology that accurately captures the physical behavior of power grid components within the simulation and optimization problems [1].

AC Power Flow (AC-PF) simulation represents the existing framework for modeling and analysis of the generated power dispatch and steady-state behavior of the power grid. Since its introduction in [2], the traditional AC-PF formulation uses variants of constant power injection representations (PQ and PV nodes) for aggregated load and generation modeling. However, these constant power abstractions, particularly PQ loads, are not voltage dependent and, therefore, represent aggregated load behavior that is in stark contrast to what is observed in the field [3]. For instance, consider the B.C. Hydro system where it was shown that decreasing the substation voltage by 1% decreased the active and reactive power demand by 1.5% and 3.4% respectively [3]. Therefore, the inaccuracies in modeling the aggregated steady-state behavior can possibly have a significant impact on the accurate simulation and optimization of the power grid steady-state response. Estimates [4] suggest that improving efficiency of power delivery could save an additional $20.4 billion per year, out of which a portion can certainly be contributed to the improvements in modeling the power grid aggregation [4].

Accurate modeling and forecasting of the power system demand has been a prominent research topic for many years [1]. A ZIP model was proposed to capture voltage magnitude dependency for improved accuracy. A ZIP model does not, however, capture the voltage angle variabilities in general [5], particularly at voltage-controlled buses where it reverts to a PQ model. Most importantly, the inaccuracies of these models and their representation of correct load voltage sensitivities is not well understood, particularly in terms of impact on optimization results.

Regarding power flow models in general, the equivalent split-circuit modeling and analysis framework has been demonstrated for improving the robustness of simulation [5]-[7] and optimization [8]. The circuit formalism that it offers further facilitated a new circuit theoretic BIG load model for capturing the behavior of any aggregated load data [5]. Of notable distinction is that the BIG model *does* characterize the voltage angle variability, in contrast to other more established ZIP and exponential load models. Furthermore, the BIG load model is linear within the current-voltage based formulation, and as with other models it can be used to fit historical or forecasted data using machine learning algorithms [5],[9]-[10].

In this paper, we utilize the circuit formalism provided by a split-circuit representation for the power grid to explore the effect of aggregated load models on the optimal power flow solution. Comparison between PQ and more complex models that capture some aspect of load impedance, hence sensitivity, is used to demonstrate that the choice of load model type can result in significantly different grid operating points at the optimal power flow solution. We introduce a new metric as an application of the maximum power transfer theorem to relating the load model sensitivities in order to examine the impact of an aggregated load model to the optimal power flow solution and further classify the applicability of load model types based on measurement data. Depending on its sensitivities (load coefficients), an aggregated load model can be classified as a power-type or an impedance-type model, and as such have a different impact on the OPF solution. The sensitivity conditions defining the power- and impedance-types for ZIP and BIG load models are derived. A similar approach can be used for the classification of any other aggregated load model.

For demonstration of the impact of the load model choice to the OPF solution, we apply a machine learning algorithm [5] to characterize the Carnegie Mellon University (CMU) campus dataset for a 2-day period in terms of PQ, ZIP and BIG load model parameters. To analyze the impact of the load model types, an AC optimal power flow analysis is performed for the CMU campus test case for the three distinct models. The results demonstrate the strong correlation between the choice of load model and the optimal power flow solution. For instance, the PQ load model accurately characterized the CMU demand in only 16.7% cases, while in all other cases it yielded significantly different voltage set points as compared to other models.

## II. BACKGROUND

### A. Optimizing power flow for the given demand

Solving the optimal power flow optimization problem represents a core component of Independent System Operator (ISO) market analysis and dispatch [11]. Namely, for a given network topology, a set forecasted load demands, and a set of committed generation units, the objective of optimal power flow problem is to determine a steady-state grid operating point that further minimizes the given cost function, i.e. $\mathcal{F}_c(X)$:

$$\min_{X} \mathcal{F}_c(X) \qquad (1)$$

while satisfying the network constraints (2) and operational limits (3) given in generic form as:

$$F_G(X) + F_N(X) + F_D(X) = \mathbf{0} \qquad (2)$$
$$F_O(X) \leq \mathbf{0} \qquad (3)$$

Where $X$ is a vector of power system state variables, and the terms from (2) correspond to the committed generation $F_G(X)$, network topology $F_N(X)$ and set of load demand $F_D(X)$ components of network constraints.

Although the definition of optimal power flow given by (1)-(3) is generic and, therefore, holds for any formulation of network and operational constraints, herein we consider the AC power grid models. Hence, the network constraint can be represented in terms of polar [2] or rectangular power mismatch formulation [11], current injection formulation [11], or the more recently proposed equivalent split-circuit formulation [8], without requiring any simplifications to the grid models. Most importantly, assuming that the network topology and committed generation units are well defined, modeling the load demand remains a possible source of introduced inaccuracy within the optimization.

### B. Modeling and forecasting the aggregated load demand

Forecasting the load aggregation has been traditionally done in terms of real and reactive power demand, while considering the SCADA data, weather forecasts and other parameters that can have an effect on power consumption [1]. However, the accurate forecasting of the demanded power does not necessarily correspond to obtaining the aggregation model that accurately captures the complete load characteristics [1],[5], such as the sensitivities due to the voltage variations. For instance, there can be many different aggregated models (PQ, ZIP, BIG, Y, etc.) that absorb the forecasted powers at a given voltage operating point.

We have recently introduced an aggregated load modeling framework that instead of characterizing and forecasting the load demand power, characterizes (PowerFit [5]) and forecasts (PowerCast [9] and StreamCast [10]) the aggregated load based on its sensitivities. Furthermore, any existing aggregated load model can be included within the framework, where, for instance, the constant PQ load model can be seen as a model with zeroth order voltage sensitivities, the ZIP model captures first order sensitivities with respect to the bus voltage magnitude, while the BIG model captures both angle and magnitude voltage sensitivities. Most importantly, any additional load sensitivities such as sensitivities to weather conditions, seasonal patterns or number of people within the area can be considered without loss of generality [9].

## III. IMPEDANCE-TYPE AND POWER-TYPE OF AN AGGREGATED LOAD MODEL

It is conceivable that two different load models can absorb the same amount of forecasted real and reactive powers at the optimal solution, while setting completely different grid operating points. Hence, the dependency between the optimal operating point of a power grid and the load modeling parameters. In this section we use a three bus test case example (Fig. 1) and circuit theory to motivate the introduction of the conditions that characterize the aggregated load model types within the power flow optimization problem.

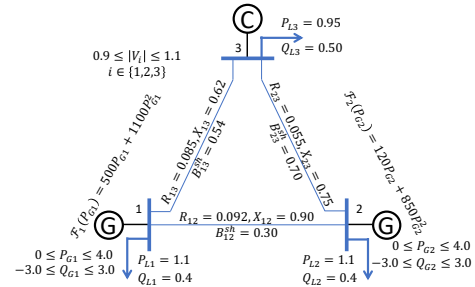

Fig. 1. Per unit parameters of the analyzed 3-bus test case.

### A. Defining a "Power-Type" load model

To introduce the power-type load model, we first examine the impact of the PQ load model on the solution space of the optimal power flow problem. Consider a complex current $(I_{RL} + jI_{IL})$ flowing into the constant PQ model, defined from the expression of complex power as:

$$I_{RL} + jI_{IL} = \frac{(P_L - jQ_L)(V_{RL} + jV_{IL})}{|V_L|^2} \qquad (4)$$

where $P_L$ and $Q_L$ represent the set constant real and reactive powers, while $|V_L|$, $V_{RL}$ and $V_{IL}$ are the load bus voltage magnitude and its real and imaginary components.

It should be noted that the load current from (4) is inversely proportional to the bus voltage magnitude. Therefore, it is implied that the gradient of the power flow optimization problem will be directed towards setting the highest feasible voltage that drives the lower current to reduce the system losses and maintain the set real and reactive power demands in a way that the generated power cost is minimized. This load model behavior can be further seen from the AC-OPF solution of the 3-bus test case presented in Fig. 2. Namely, the system operating point is set by the upper voltage magnitude bounds that further draw the lowest currents constrained by the constant PQ loads in order to minimize the transmission losses, hence the cost of generated real power.

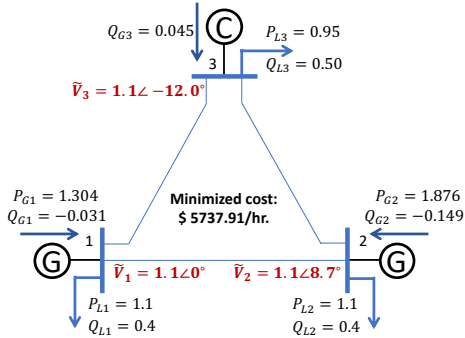

Fig. 2. AC-OPF p.u. solution of the 3-bus test case with PQ loads.

Next, let $P_{AL}(V)$ and $Q_{AL}(V)$ be the real and reactive powers absorbed by a load bus, further dependent on a voltage state variable vector, $V$. We generalize the recognized load model behavior within the power flow optimization space, by defining the "power-type" of a load model:

**Definition 1** (Power type of an aggregated load model). *An aggregated load bus model whose sensitivities ensure that the power supplied to the load for positive voltage perturbation is equal or less than the power absorbed at a nominal point. Therefore, the optimal solution seeks the highest feasible voltage operating point to minimize the total generation cost. This relationship can be further expressed in terms of real and reactive power sensitivity conditions as:*

$$\frac{\partial P_{AL}(V)}{\partial V} \leq 0, \frac{\partial Q_{AL}(V)}{\partial V} \leq 0 \quad (5)$$

### B. Defining an "Impedance-Type" load model

Let $P_Y$ and $Q_Y$ be the forecasted real and reactive power demanded at a bus with predominantly impedance aggregation. Hence, the further representation of the load bus in terms of a constant power model within the power flow optimization problem, can significantly affect the optimal operating point, and therefore, introduce inaccuracies in further analysis. Herein, we demonstrate that the relationship between the forecasted power demand and a load aggregated model is not unique such that it can be generally represented by a constant power model, by introducing an "impedance-type" aggregated load model.

In contrast to the constant power relationship from (4), the current and voltage related by an impedance are directly proportional, as defined by the Ohm's Law. Therefore, it can be inferred that the gradient of the power flow optimization problem goes toward finding the lowest feasible operating point that corresponds to the lower currents that reduce the system losses, thereby minimizing the generated power cost. To demonstrate this behavior, consider the 3-bus test case from Fig. 3, with the constant PQ load models replaced by equivalent impedances. As can be seen, even though the loads absorb the same amount of real and reactive power, the voltage operating point changes significantly from that in Fig. 2.

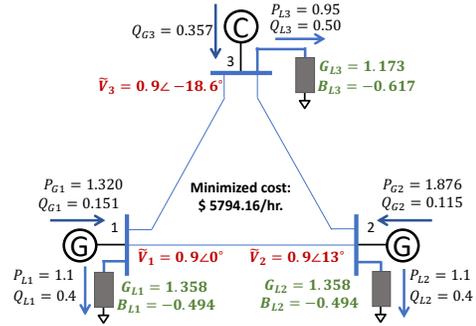

Fig. 3. AC-OPF p.u. solution of the 3-bus test case with Y loads.

We can generalize an impedance-type aggregated load as:

**Definition 2** (Impedance-type of an aggregated load model). *Type of a load bus model for which sensitivities ensure that the power supplied to the load for positive voltage perturbation is greater than the power absorbed at a nominal point. Therefore, the optimal solution seeks the lowest feasible bus voltage operating point to minimize the total generation cost. This relationship can be further expressed in terms of real and reactive power sensitivity conditions as:*

$$\frac{\partial P_{AL}(V)}{\partial V} > 0, \frac{\partial Q_{AL}(V)}{\partial V} > 0 \quad (6)$$

With these two definitions for aggregated load model types, we next apply the established conditions to show how the ZIP and BIG load model sensitivities define their behavior within the power flow optimization space.

## IV. CLASSIFICATION OF ZIP AND BIG LOAD MODELS

### A. ZIP load model

Introduced as an improvement over the constant power model, a voltage dependent ZIP load model captures the load demand sensitivities with respective to bus voltage magnitude. The total power absorbed by the ZIP load can be written as a combination of powers set by the constant power load with additions of powers absorbed by the current and impedance segments of the load as:

$$P_{ZIP} = P_0 + I_P|V_L| + G_Z|V_L|^2 \quad (7)$$
$$Q_{ZIP} = Q_0 + I_Q|V_L| + B_Z|V_L|^2 \quad (8)$$

where $P_0$ and $Q_0$ are the constant power terms, and $I_P$ $I_Q$, $G_{ZIP}$, and $B_{ZIP}$ represent the current and impedance terms that act like sensitivities around the base constant powers.

To explore the behavior of a ZIP load model within an OPF problem formulation, we apply the postulated conditions from Definitions 1 and 2 to the ZIP model governing equations from (7)-(8). For instance, given a real power cost minimization problem, the ZIP load model represents a power type load if the real power conditions from (5) are satisfied:

$$\frac{\partial P_{ZIP}}{\partial |V_L|} \leq 0 \Rightarrow I_P + 2G_Z|V_L| \leq 0 \quad (9)$$

Note that the same condition can be obtained from the sensitivities of real and imaginary voltage components.

Interestingly, if we further multiply the condition from (9) by the voltage magnitude variable, we obtain the condition which implies that *a real power supplied by the current portion of a power-type ZIP load model has to be equal or less than twice the real power absorbed by the conductance portion of the load.* Conversely, if this condition is not true, then the ZIP load model represents an impedance type model.

### B. BIG load model

The circuit theoretic BIG load model [5][9]-[10] was recently introduced as an alternative to the ZIP model that captures both voltage magnitude and angle sensitivities. In contrast to ZIP, the BIG model was shown to be linear within the current injection and equivalent circuit formulations, while preserving the accuracy in characterizing the aggregated load behavior [5]. Most importantly, the sensitivities of a BIG load model further represent the small signal model of an aggregated load compatible with time-domain simulations.

The governing equations of a BIG load model can be written in terms of real and imaginary currents as:

$$I_{R,BIG} = \alpha_R + G_B V_{RL} - B_B V_{IL} \quad (10)$$
$$I_{I,BIG} = \alpha_I + G_B V_{IL} + B_B V_{RL} \quad (11)$$

where, $\alpha_R$ and $\alpha_I$ are the constant current parameters, while $G_{BIG}$ and $B_{BIG}$, represent the conductance and susceptance sensitivity elements respectively.

By considering the same steps as in derivation of classifying conditions for ZIP load model, by Definition 1, BIG represents a power-type load model when the following two conditions are satisfied:

$$\frac{\partial P_{BIG}}{\partial V_{RL}} \leq 0 \Longrightarrow \alpha_R + 2G_B V_{RL} \leq 0$$
$$\frac{\partial P_{BIG}}{\partial V_{IL}} \leq 0 \Longrightarrow \alpha_I + 2G_B V_{IL} \leq 0 \quad (12)$$

It should be noted that the same conditions hold if the differentiation of power is obtained with respect to the voltage magnitude and angle variables. In that case, it can be shown that, the real power supplied by the current source ($\alpha_R + j\alpha_I$) of a power-type BIG load model has to be equal or less than the twice the real power absorbed by the conductance portion of the load, as in the case of ZIP model.

Most importantly, as it can be recognized from the case of ZIP and BIG models, *the conditions defining the aggregated load model types represent the application of maximum power transfer theory to the parameters of load model.* Namely, in a power-type aggregated load model, the power absorbed by the voltage sensitivity elements is greater than half of the power supplied by the other model terms, otherwise, the aggregated load model represents an admittance-type.

## V. SIMULATION RESULTS

In this section we verify the postulated load model classification based on the voltage sensitivities, and further analyze its impact on the optimal power flow solution of a realistic test case. Our PowerFit algorithm [5] was used to characterize the normalized measurement data obtained for a two-day period at the Carnegie Mellon University (CMU) campus in terms of PQ, ZIP and BIG load models. To present a meaningful comparison between the models, PowerFit found the six daily segments that yielded similar average RMS characterization errors (~2-3%) for all three model types. For instance, the PowerFit input data for the real current measurement of the University Center current is shown fitted to the three different models in Fig. 4. Each segment represents a change in model parameters needed to fit the time series data with the accuracy shown.

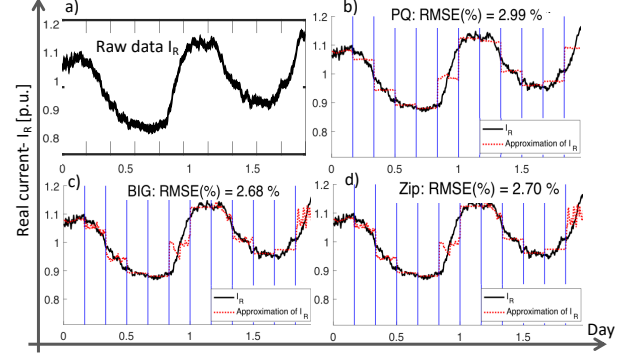

Fig. 4. Univ. Center (L4) real current measurement - CMU campus dataset.

The load model coefficients for the 12 segments found by PowerFit were incorporated into a 5-bus representation of the CMU campus (Fig. 5). The PQ, ZIP and BIG parameters from a first segment are presented in Tables 1-3, while the complete set of load parameters can be found in www.andrew.cmu.edu/user/bhooi/loadmodels/CMU_campus_test_case_data.zip.

Table 1: PQ load parameters for the first segment of CMU campus data.

| $P_{L2}^1$ [p.u.] | $Q_{L2}^1$ [p.u.] | $P_{L3}^1$ [p.u.] | $Q_{L3}^1$ [p.u.] | $P_{L4}^1$ [p.u.] | $Q_{L4}^1$ [p.u.] |
|---|---|---|---|---|---|
| 1.4499 | 0.44594 | 2.0868 | 0.64185 | 1.0589 | 0.32567 |

Table 2: ZIP load parameters for the first segment of CMU campus data.

| Load bus | $G_Z^1$ [p.u.] | $B_Z^1$ [p.u.] | $P_0^1$ [p.u.] | $Q_0^1$ [p.u.] | $I_P^1$ [p.u.] | $I_Q^1$ [p.u.] |
|---|---|---|---|---|---|---|
| L2 | -0.16338 | -0.50372 | 1.1392 | -0.19632 | 0.4767 | 0.15877 |
| L3 | 0.42845 | -0.36053 | 0.98408 | 0.076938 | 0.70154 | 0.22081 |
| L4 | 0.21739 | -0.18293 | 0.49932 | 0.03904 | 0.35596 | 0.11204 |

Table 3: BIG load parameters for the first segment of CMU campus data.

| Load bus | $G_B$ [p.u.] | $B_B$ [p.u.] | $\alpha_R$ [p.u.] | $\alpha_I$ [p.u.] |
|---|---|---|---|---|
| L2 | -0.10372 | -0.49365 | 1.5775 | 0.031136 |
| L3 | -0.14931 | -0.7106 | 2.2709 | 0.044819 |
| L4 | 0.43166 | -0.19968 | 0.65358 | -0.13531 |

AC-OPF is performed for each of the 12 segments and load models on the CMU test case. The respective cost and voltage operating point comparisons are shown in Fig. 6 and Table 4.

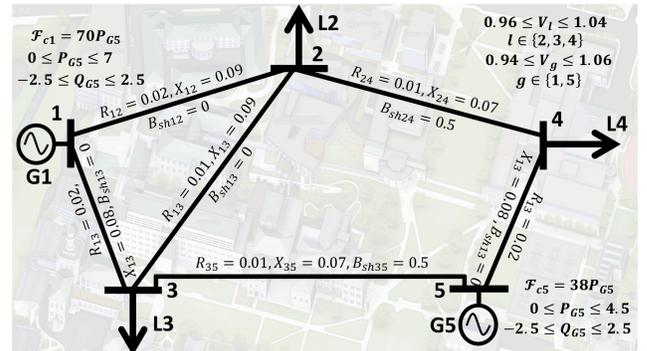

Fig. 5. Five bus test case representation of a CMU campus. Note that all of the parameters are in given their p.u. values.

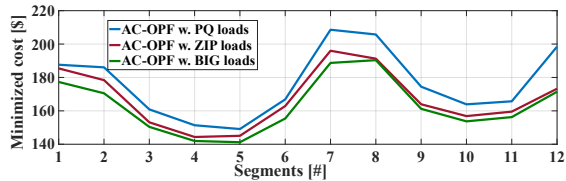

Fig. 6. Minimized generation cost comparison for the CMU test case.

Table 4: Voltage magnitude operating point comparisons for CMU test case

| Segment: | | #1 | #2 | #3 | #4 | #5 | #6 | #7 | #8 | #9 | #10 | #11 | #12 |
|---|---|---|---|---|---|---|---|---|---|---|---|---|---|
| Bus 1 | PQ | 1.04 | 1.04 | 1.038 | 1.038 | 1.039 | 1.038 | 1.044 | 1.043 | 1.038 | 1.038 | 1.038 | 1.042 |
| | ZIP | 0.958 | 1.054 | 0.974 | 0.963 | 0.968 | 0.94 | 0.966 | 1.038 | 0.953 | 0.954 | 0.972 | 0.94 |
| | BIG | 0.994 | 1.015 | 0.992 | 0.99 | 0.99 | 0.991 | 0.999 | 0.999 | 0.997 | 0.994 | 0.995 | 0.998 |
| Bus 2 | PQ | 1.01 | 1.007 | 1.011 | 1.014 | 1.015 | 1.01 | 1.006 | 1.005 | 1.006 | 1.009 | 1.008 | 1.004 |
| | ZIP | 0.974 | 1.039 | 0.964 | 0.962 | 0.96 | 1.036 | 0.96 | 1.04 | 0.965 | 0.962 | 0.96 | 0.983 |
| | BIG | 0.96 | 1.01 | 0.96 | 0.96 | 0.96 | 0.96 | 0.96 | 1.008 | 0.96 | 0.96 | 0.96 | 0.96 |
| Bus 3 | PQ | 1.016 | 1.013 | 1.016 | 1.019 | 1.02 | 1.016 | 1.012 | 1.012 | 1.012 | 1.015 | 1.014 | 1.011 |
| | ZIP | 0.96 | 1.04 | 0.96 | 0.96 | 0.962 | 1.005 | 0.972 | 1.04 | 0.96 | 0.96 | 0.962 | 0.96 |
| | BIG | 0.965 | 1.006 | 0.965 | 0.964 | 0.964 | 0.963 | 0.966 | 1.01 | 0.965 | 0.965 | 0.965 | 0.967 |
| Bus 4 | PQ | 1.012 | 1.009 | 1.014 | 1.018 | 1.019 | 1.013 | 1.008 | 1.007 | 1.009 | 1.012 | 1.011 | 1.007 |
| | ZIP | 0.962 | 1.036 | 0.966 | 0.965 | 0.96 | 1.012 | 0.96 | 1.034 | 0.963 | 0.96 | 0.96 | 0.965 |
| | BIG | 0.96 | 1.002 | 0.96 | 0.96 | 0.96 | 0.96 | 0.96 | 1.00 | 0.96 | 0.96 | 0.96 | 0.96 |
| Bus 5 | PQ | 1.06 | 1.06 | 1.06 | 1.06 | 1.06 | 1.06 | 1.06 | 1.06 | 1.06 | 1.06 | 1.06 | 1.06 |
| | ZIP | 0.971 | 1.06 | 0.972 | 0.963 | 0.981 | 0.94 | 0.994 | 1.05 | 0.984 | 0.986 | 0.989 | 0.953 |
| | BIG | 1.012 | 1.025 | 1.004 | 1.001 | 1.00 | 1.002 | 1.014 | 1.017 | 1.008 | 1.006 | 1.007 | 1.014 |

| Load bus exhibiting characteristic P-behavior | Load exhibiting characteristic Y-behavior |
| Produced different optimal generator voltage setpoint as compared to that from ZIP/BIG models | |

As expected, and as shown in Fig. 6, the minimized costs follow the load demand change (Fig. 4) and are vary with each of the load model types. The PQ loads results in the highest cost due to the inability to capture the voltage sensitivities, which also corresponds to the conservative nature of the model. The ZIP model captures the voltage sensitivities and, therefore, has a lower cost. However, ZIP model cost is higher than the one obtained with the BIG models due to the conservative constant power term in the ZIP model.

Table 4 presents the optimal voltage magnitude setpoints obtained from AC-OPF analysis with PQ, BIG and ZIP load representations for the 12 segments in Fig. 4. From Table 4 we recognize some significant deviations (*beige shadings*) in bus voltages obtained from the AC-OPF with voltage dependent load models versus that with PQ models, particularly at the generator buses (buses 1 and 5). Importantly, this indicates the limitations of using the PQ load model which less accurately characterized the CMU load behavior to run the optimal power flow analysis. These results further imply that using the PQ-load optimal solution to set the grid voltages can conceivably lead to increased generation power that is conservative. Interestingly, the voltage sensitive CMU load representations are mostly characterized by the admittance-type models (*green shadings*) and exhibit the power-type behavior (*blue shadings*) only at a single segment that repeated in a daily pattern. Note that these are the only cases for which the set points for buses 1 and 5 based on the PQ model match those based on the ZIP and BIG model, since the load is behaving as power-type.

As a final experiment, we used the PQ load models to run the OPF and determine the optimal set points for the power system defined by ZIP loads. The power flow simulation of these PQ-based OPF set points performed with the ZIP load models shows that additional generated power is needed for the CMU campus test case as a direct consequence of using inaccurate load models for the optimization problem. The results are shown in Fig. 7.

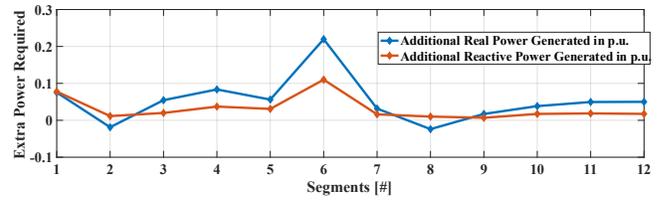

Fig. 7. Additional power required due to the use of less accurate load model.

As implied from the simulation results, representing the load models inaccurately can significantly affect the optimal system operating point. These results raise questions regarding the validity of an AC-OPF solution defined with PQ load models.

## VI. CONCLUSIONS

In this paper we apply circuit theory to analyze the impact of load models on the optimal power flow solution. It is shown that based on the maximum power transfer theorem, a load model can be classified as power type or impedance type. These load model classifications are analyzed for a CMU campus dataset. The preliminary results confirm the postulated correlation between the load model types and the optimal grid operating point. Our analyses show that replacing PQ load models with ZIP or BIG models provides more accurate and realistic power grid solutions. Furthermore, the BIG model offers load model linearity within the current injection and equivalent circuit formulations and represents a step toward the unification of power flow and time domain analyses.

ACKNOWLEDGMENT

This work was supported in part by the Defense Advanced Research Projects Agency (DARPA) under award no. FA8750-17-1-0059 for RADICS program, and the National Science Foundation (NSF) under contract no. ECCS-1800812.